# Dirac bound state solutions of spherically ring-shaped $q$-deformed Woods-Saxon potential for any $l$-state

Sameer M. Ikhdair[1*], Majid Hamzavi[2**]
[1]Physics Department, Near East University, Nicosia, North Cyprus, Mersin 10, Turkey
[2]Department of Basic Sciences, Shahrood Branch, Islamic Azad University, Shahrood, Iran
* sikhdair@neu.edu.tr
** Corresponding Author: majid.hamzavi@gmail.com

## Abstract

Approximate bound state solutions of the Dirac equation with $q$-deformed Woods-Saxon plus a new generalized ring-shaped potential are obtained for any arbitrary $l$-state. The energy eigenvalue equation and corresponding two-component wave function are calculated by solving the radial and angular wave equations within a shortcut of the Nikiforov-Uvarov method. The solutions of the radial and polar angular parts of the wave function are expressed in terms of the Jacobi polynomials. A new approximation being expressed in terms of the potential parameters is carried out to deal with the strong singular centrifugal potential term $l(l+1)r^{-2}$. Under some limitations, we can obtain solution for the ring-shaped Hulthén potential and the standard usual spherical Woods-Saxon potential ($q=1$).

**Keywords:** Dirac equation, $q$-deformed Woods-Saxon potential, ring-shaped potential, approximation scheme, Nikiforov-Uvarov method
**PACS:** 03.65.Ge, 03.65.Fd, 0.65.Pm, 02.30.Gp

## 1- Introduction

The Woods-Saxon (WS) potential is a realistic short range potential and used to study the nuclear structure within the shell model in nuclear physics [1]. Different versions of this potential have been introduced to explore elastic and quasi-elastic scattering of nuclear particles [2]. The usual ($q=1$) and the $q$-deformed WS potentials have been applied in nuclear calculations [3-11], studying the behavior of valence electrons in metallic systems or in helium model [12] and in nonlinear scalar theory of mesons [13].

The Schrödinger equation [14-17], Klein-Gordon equation [18-20] and Dirac equation [21] have been solved with the usual and deformed WS potential forms for their bound states in the framework of the Nikiforov-Uvarov method and by other methods [22-24].

Recently, a new ring-shaped (RS) potential has been introduced plus Coulomb potential [25], Hulthen potential [26], modified Kratzer potential [27] and nonharmonic oscillator potential [28]. Such calculations with this RS potential have found applications in quantum chemistry such as the study of ring-shaped molecules like benzene. Overmore, the shape form of this potential play an important role when studying the structure of deformed nuclei or the nuclear interactions. Quesne [29] obtained a new RS potential by replacing the Coulomb part of Hartmann potential [30] by a harmonic oscillator term. Gang [31] exactly obtained energy spectrum of some non-central separable potential in $r$ and $\theta$ using the method of supersymmetric WKB approximation. Yasuk *et al* [32] obtained general solutions of Schrödinger equation for a non-central potential by using the Nikiforov-Uvarov (NU) method [33]. Yuan *et al* [34] studied exact solutions of scattering states of the Klein-Gordon equation with Coulomb potential plus new RS potential with equal mixture of scalar and vector potentials. Ikhdair and Sever [35] used the polynomial solution to solve a non-central potential. Gang and Bang [36] studied the Klein-Gordon with equal scalar and vector Makarov potentials by the factorization method. Kerimov [37] studied non-relativistic quantum scattering problem for a non-central potential which belongs to a class of potentials exhibiting an ‘accidental degeneracy. Berkdemir and Sever [38] investigated the diatomic molecules subject to central potential plus ring-shaped potential. Also they [39] solved the pseudospin symmetric solution of the Dirac equation for spin-$1/2$ particles moving with the Kratzer potential connected with an angle-dependent potential systematically. Yeşiltaş [40] showed that a wide class of non-central potentials can be analyzed via the improved picture of the NU method. Berkdemir and Cheng [41] investigated the problem of relativistic motion of a spin-$1/2$ particle in an exactly solvable potential consisting of harmonic oscillator potential plus a novel ring-shaped dependent potential. Zhang *et al* [42-44] obtained the complete solutions of the Schrödinger and Dirac equations with a spherically harmonic oscillatory ring-shaped potential. Ikhdair and Sever obtained the exact solutions of the D-dimensional Schrödinger equation with ring-shaped pseudoharmonic potential [45], modified Kratzer potential [46] and the D-dimensional Klein-Gordon equation with ring-shaped pseudoharmonic potential [47]. Hamzavi *et al* found the exact solutions of Dirac equation with Hartmann potential [48] and ring-shaped pseudoharmonic oscillatory potential [49] by using NU method. Many authors have also studied a few non central potentials within the supersymmetric quantum mechanics and point canonical transformations [50-52].

The aim of this work is to investigate analytical bound-state solutions of the Dirac equation with non central $q$-deformed WS potential plus a new generalized RS potential with extra additional parameter $\alpha$ from the RS potential used in [25]. Therefore, the non central potential of the type $V(\vec{r}) = V_{WS}(r) + \frac{1}{r^2}V_{RS}(\theta)$, consisting of two parts:

$$V_{WS}(r) = -\frac{V_0}{1+qe^{(r-R_0)/a}}, \; V_{RS}(\theta) = \frac{\alpha+\beta\cos^2\theta}{\sin^2\theta}, \tag{1}$$

with $V_{WS}(r)$ is the $q$-deformed WS potential in which $V_0$, $R_0$, $a$ and $q$ are the potential depth, width or nuclear radius, surface thickness and deformation parameters, respectively. Further, $V_{RS}(\theta)$ is a new RS potential identical to the RS part of the Hartmann potential [25]. Here $\alpha = -p\sigma^2\eta^2 a_0^2 \varepsilon_0$ and $\beta = -p\sigma^2\eta^2 a_0^2 \varepsilon_0$, where $a_0 = \frac{\hbar^2}{me^2}$ and $\varepsilon_0 = -\frac{me^4}{2\hbar^2}$ represent the Bohr radius and the ground-state energy of the hydrogen atom, respectively. Further, $\eta$, $\sigma$ and $p$ are three dimensionless parameters. Generally speaking, $\eta$ and $\sigma$ vary from about $1$ up to $10$ and $p$ is a real parameter and its value is taken as $1$.

We also show that when the deformation parameter $q$ takes a particular value, the results turn to be the solution for the Hulthen potential.

In our solution, we are using a powerful shortcut of the NU method [33] that has proven its efficiency and easily handling in the treatment of problems with second-order differential equations of the type $y'' + (\tilde{\tau}/\sigma)y' + (\tilde{\sigma}/\sigma^2)y = 0$ which are usually encountered in physics such as the radial and angular parts of the Schrödinger, KG and Dirac equations [45-49].

This paper is organized as follows. In section 2, we present the Dirac equation for the generalized ring-shaped $q$-deformed WS potential. Section 3 is devoted to derive the approximate analytic bound-state energy eigenvalue equation and the associated two-components of the wave function consisting from radial and angular parts within a shortcut of the NU method. Section 4 presents the conclusion of our work.

## 2. Dirac equation with scalar and vector $q$-deformed WS plus ring-shaped potential

The Dirac equation for a particle of mass $M$ moving in the field of attractive scalar potential $S(\vec{r})$ and repulsive vector potential $V(\vec{r})$ potentials (in the relativistic units $\hbar = c = 1$) takes the form [53]

$$\left[\vec{\alpha}\cdot\vec{p}+\beta\left(M+S(r)\right)+V(r)\right]\psi(\vec{r})=E\psi(\vec{r}), \tag{2}$$

with $E$ is the relativistic energy of the system and $\vec{p}=-i\vec{\nabla}$ is the three-dimensional (3D) momentum operator. Further, $\vec{\alpha}$ and $\beta$ represent the $4\times 4$ usual Dirac matrices given by

$$\vec{\alpha}=\begin{pmatrix}0 & \sigma_i\\ \sigma_i & 0\end{pmatrix},\quad \beta=\begin{pmatrix}I & 0\\ 0 & -I\end{pmatrix},\qquad i=1,2,3 \tag{3}$$

which are expressed in terms of the three $2\times 2$ Pauli matrices

$$\sigma_1=\begin{pmatrix}0 & 1\\ 1 & 0\end{pmatrix},\ \sigma_2=\begin{pmatrix}0 & -i\\ i & 0\end{pmatrix},\ \sigma_3=\begin{pmatrix}1 & 0\\ 0 & -1\end{pmatrix}, \tag{4}$$

and $I$ is the $2\times 2$ unitary matrix. In addition, the Dirac wave function $\psi(\vec{r})$ can be expressed in Pauli-Dirac representation as

$$\psi(\vec{r})=\begin{pmatrix}\varphi(\vec{r})\\ \chi(\vec{r})\end{pmatrix}. \tag{5}$$

Inserting Eqs. (3) to (5) into Eq. (2) give

$$\vec{\sigma}\cdot\vec{p}\chi(\vec{r})=\left(E-M-\Sigma(\vec{r})\right)\varphi(\vec{r}), \tag{6a}$$

$$\vec{\sigma}\cdot\vec{p}\varphi(\vec{r})=\left(E+M-\Delta(\vec{r})\right)\chi(\vec{r}), \tag{6b}$$

where the sum and difference potentials, respectively, are defined by

$$\Sigma(\vec{r})=V(\vec{r})+S(\vec{r})\ \text{ and }\ \Delta(\vec{r})=V(\vec{r})-S(\vec{r}). \tag{7}$$

For a limiting case when $S(\vec{r}) = V(\vec{r}),$ then $\Sigma(\vec{r}) = 2V(\vec{r})$ and $\Delta(\vec{r}) = 0.$ Consequently, Eq. (6) becomes

$$\vec{\sigma} \cdot \vec{p}\chi(\vec{r}) = \left(E - M - 2V(\vec{r})\right)\varphi(\vec{r}), \tag{8a}$$

$$\chi(\vec{r}) = \frac{\vec{\sigma} \cdot \vec{p}}{E + M}\varphi(\vec{r}), \tag{8b}$$

where $E \neq -M$, which means that only the positive energy states do exist for a finite lower-component $\chi(\vec{r})$ of the wave function.

Combining Eq. (8b) into Eq. (8a) and inserting the potential (1), one can obtain

$$\left[\nabla^2 + E^2 - M^2 + 2\left(E + M\right)\left(\frac{V_0}{1 + qe^{(r-R_0)/a}} - \frac{\alpha + \beta\cos^2\theta}{r^2\sin^2\theta}\right)\right]\varphi_{nlm}(r,\theta,\phi) = 0. \tag{9}$$

where

$$\nabla^2 = \frac{1}{r^2}\left[\frac{\partial}{\partial r}\left(r^2\frac{\partial}{\partial r}\right) + \frac{1}{\sin\theta}\frac{\partial}{\partial\theta}\left(\sin\theta\frac{\partial}{\partial\theta}\right) + \frac{1}{\sin^2\theta}\frac{\partial^2}{\partial\phi^2}\right]. \tag{10}$$

and

$$\varphi_{nlm}(r,\theta,\phi) = R_{nl}(r)Y_l^m(\theta,\phi),\ R_{nl}(r) = r^{-1}U_{nl}(r),\ Y_l^m(\theta,\phi) = \Theta_l(\theta)\Phi_m(\phi). \tag{11}$$

After substituting Eqs. (10) and (11) into Eq. (9) and making a separation of variables, we obtain the following sets of second-order differential equations:

$$\frac{d^2U_{nl}(r)}{dr^2} + \left[E^2 - M^2 - \frac{\lambda}{r^2} + \frac{2\left(E + M\right)V_0}{1 + qe^{(r-R_0)/a}}\right]U_{nl}(r) = 0, \tag{12a}$$

$$\frac{d^2\Theta_l(\theta)}{d\theta^2} + \cot\theta\frac{d\Theta_l(\theta)}{d\theta} + \left[\lambda - \frac{m^2}{\sin^2\theta} - \frac{2\left(E + M\right)\left(\alpha + \beta\cos^2\theta\right)}{\sin^2\theta}\right]\Theta_l(\theta) = 0, \tag{12b}$$

$$\frac{d^2\Phi_m(\phi)}{d\phi^2} + m^2\Phi_m(\phi) = 0, \tag{12c}$$

where $m^2$ and $\lambda = l(l+1)$ are two separation constants with $l$ is the rotational angular momentum quantum number.

The solution of Eq. (12c) is periodic and must satisfy the periodic boundary condition

$$\Phi_m(\phi + 2\pi) = \Phi_m(\phi), \tag{13}$$

which gives the solution:

$$\Phi_m(\phi) = \frac{1}{\sqrt{2\pi}}\exp\left(\pm im\phi\right), \qquad m = 0,1,2,\cdots. \tag{14}$$

The solutions of the radial part (12a) and polar angular part (12b) equations will be shown in the later section.

## 3. Analytical solutions of the radial and polar angle parts of Dirac equation

### 3.1. Solution of polar angle part

To obtain the energy eigenvalues and eigenfunctions of the polar angular part of Dirac equation (12b), we make an appropriate transformation of parameter as $z=\cos^2\theta$ (or $z=\sin^2\theta$) to reduce it as

$$\Theta_l''(z)+\left[\frac{(1/2)-(3/2)z}{z(1-z)}\right]\Theta_l'(z)+\frac{1}{z^2(1-z)^2}$$

$$\times\left[-\frac{1}{4}[\lambda+2(E+M)\beta]z^2+\left[\lambda-m^2-2(E+M)\alpha\right]z\right]\Theta_l(z)=0, \quad (15)$$

where $\Theta_l(z=0)=0$ and $\Theta_l(z=1)=0.$ The solution of the above angular equation can be easily found by using the shortcut of the NU method presented in Appendix A. Now, in comparing the above equation with Eq. (A2), we identify the following constants:

$$c_1=\frac{1}{2},\ c_2=\frac{3}{2},\ c_3=1,\ A=\frac{1}{4}[\lambda+2(E+M)\beta],\ B=\left[\lambda-m^2-2(E+M)\alpha\right],\ C=0.$$

The remaining constants are thus calculated via (A5) as

$$c_4=\frac{1}{4},\ c_5=-\frac{1}{4},\ c_6=\frac{1+4\lambda+8(E+M)\beta}{4},\ c_7=\frac{m^2-\lambda+2(E+M)\alpha}{4}-\frac{1}{8},\ c_8=\frac{1}{16},$$

$$c_9=\frac{m^2+2(E+M)(\alpha+\beta)}{4},\ c_{10}=\frac{1}{2},\ c_{11}=\sqrt{m^2+2(E+M)(\alpha+\beta)},\ c_{12}=\frac{1}{2},$$

$$c_{13}=\frac{1}{2}\sqrt{m^2+2(E+M)(\alpha+\beta)}. \quad (16)$$

We use the energy relation (A10) and the parametric coefficients given by Eqs. (15) and (16) to obtain a relationship between the separation constant $\lambda$ and the new non negative angular integer $\tilde{n}$ as

$$\lambda=l(l+1)=\left(2\tilde{n}+\sqrt{m^2+2(E+M)(\alpha+\beta)}+\frac{3}{2}\right)^2-2(E+M)\beta-\frac{1}{4}, \quad (17a)$$

$$l=\sqrt{\left(2\tilde{n}+\sqrt{m^2+2(E+M)(\alpha+\beta)}+\frac{3}{2}\right)^2-2(E+M)\beta-\frac{1}{2}}. \quad (17b)$$

Once the ring-shaped potential is disappeared after setting the potential parameters to zero, i.e., $\alpha=\beta=0$ or simply the angular part $V_{RS}(\theta)=0,$ we obtain $l=2\tilde{n}+|m|+1,\ m=0,1,2,\cdots.$ The angular part of the potential (1), $V_{RS}(\theta),$ is found to have singularities at angles $\theta = P\pi \quad (P = 0,1,2,3,\cdots)$ as well as at very small and very large values of $r$.

Let us find the corresponding polar angular part of the wave function. We find the weight function via (A11) as

$$\rho(z) = z^{1/2}\left(1-z\right)^{\sqrt{m^2+2(E+M)(\alpha+\beta)}}, \tag{18}$$

which gives the first part of the angular wave function via (A13) in terms of the Jacobi polynomial as

$$y_{\tilde{n}}(z) \sim P_{\tilde{n}}^{\left(1/2,\sqrt{m^2+2(E+M)(\alpha+\beta)}\right)}\left(1-2z\right). \tag{19}$$

The second part of the angular wave function can be obtained via (A12) as

$$\phi(z) \sim z^{1/2}\left(1-z\right)^{\sqrt{m^2+2(E+M)(\alpha+\beta)}/2}, \tag{20}$$

and hence the angular part of the wave function can be obtained via Eq. (A14); namely, $\Theta_l(z) = \phi(z)y_{\tilde{n}}(z)$ as

$$\Theta_l(\theta) = A_{\tilde{n}}\cos\theta\left(\sin\theta\right)^{\sqrt{m^2+2(E+M)(\alpha+\beta)}}\,P_{\tilde{n}}^{\left(1/2,\sqrt{m^2+2(E+M)(\alpha+\beta)}\right)}\left(1-2\cos^2\theta\right), \tag{21}$$

where $A_{\tilde{n}}$ is the normalization factor. When the ring-shaped potential is disappeared, i.e., $\alpha = \beta = 0$, then

$$\Theta_l(\theta) = A_{\tilde{n}}\cos\theta\sin^{|m|}\theta P_{\tilde{n}}^{(1/2,|m|)}\left(1-2\cos^2\theta\right).$$

### 3.2. Solution of radial part equation

In this part we will consider the energy eigenvalue equation and the wave function of the radial part of the Dirac equation with the $q$-deformed WS potential. The exact solution is not handy due to existence of the strong singular centrifugal potential term $\lambda r^{-2}$ in Eq. (12a). Therefore, an approximate analytical solution has been done for this term in Appendix B by using the following change of variable, $x = (r-R_0)/R_0$ and $\gamma = R_0/a$. Thus, Eq. (12a) becomes

$$U''_{nl}(x) + \left[R_0^2\left(E^2-M^2\right) - \lambda\left(D_0 + D_1 v + D_2 v^2\right) + 2\left(E+M\right)V_0 R_0^2 v\right]U_{nl}(x) = 0,$$

$$v = \frac{1}{1+q\,e^{\gamma x}}, \tag{22}$$

where the explicit forms of the constants $D_i$ $(i=0,1,2)$ are derived explicitly in Appendix B. Furthermore, making a change of variables as $s=e^{\gamma x}$, we can recast Eq. (22) into the simple form

$$U''_{nl}(s)+\frac{(1+qs)}{s(1+qs)}U'_{nl}(s)+\frac{\left[-\tilde{E}q^2s^2+\left(\delta-2\tilde{E}\right)qs-\left(\tilde{E}+\eta-\delta\right)\right]}{s^2\left(1+qs\right)^2}U_{nl}(s)=0, \tag{23}$$

with

$$\tilde{E}=\frac{\lambda D_0-R_0^2\left(E^2-M^2\right)}{\gamma^2},\quad \delta=\frac{2(E+M)V_0R_0^2-\lambda D_1}{\gamma^2},\quad \eta=\frac{\lambda D_2}{\gamma^2}. \tag{24}$$

Comparing Eq. (23) with its counterpart hypergeometric equation (A2), we identify values of the following constants:

$$c_1=1,\ c_2=-q,\ c_3=-q,\ A=q^2\tilde{E},\ \ B=q\left(\delta-2\tilde{E}\right),\ \ C=\tilde{E}+\eta-\delta, \tag{25}$$

and the remaining constants are calculated via (A5) as

$$c_4=0,\ c_5=\frac{q}{2},\ c_6=\frac{q^2}{4}\left(1+4\tilde{E}\right),\ c_7=q\left(2\tilde{E}-\delta\right),\ c_8=\tilde{E}+\eta-\delta,$$

$$c_9=\frac{q^2}{4}(1+4\eta),\ c_{10}=2\sqrt{\tilde{E}+\eta-\delta},\ \ c_{11}=-\frac{1}{q}\sqrt{q^2\left(1+4\eta\right)},\ c_{12}=\sqrt{\tilde{E}+\eta-\delta},$$

$$c_{13}=\frac{1}{2}\left(1-\frac{1}{q}\sqrt{q^2\left(1+4\eta\right)}\right). \tag{26}$$

The energy eigenvalue equation can be obtained via the relation (A10) and values of constants given by Eq. (25) and Eq. (26) after lengthy but straightforward algebra as

$$E^2-M^2+2(E+M)V_0=\frac{l(l+1)}{R_0^2}(D_0+D_1+D_2)$$

$$-\frac{a^2}{4}\left[\frac{2(E+M)V_0-l(l+1)(D_1+D_2)/R_0^2}{n+\frac{1}{2}-\frac{1}{2q}\sqrt{q^2\left(1+\frac{4l(l+1)D_2}{a^2R_0^2}\right)}}-\frac{1}{a^2}\left(n+\frac{1}{2}-\frac{1}{2q}\sqrt{q^2\left(1+\frac{4l(l+1)D_2}{a^2R_0^2}\right)}\right)\right]^2, \tag{27}$$

and recalling that $l(l+1)=\left(2\tilde{n}+\sqrt{m^2+2(E+M)(\alpha+\beta)}+\frac{3}{2}\right)^2-2(E+M)\beta-\frac{1}{4}.$

On the other hand, in the nonrelativistic limiting case where $E+M\simeq 2M$ and $E-M\simeq E$, Eq. (27) becomes

$$E_{nl'm}=-V_0+\frac{l'(l'+1)}{2MR_0^2}\left(D_0+D_1+D_2\right)$$

$$-\frac{a^2}{8M}\left[\frac{4MV_0-l'(l'+1)(D_1+D_2)/R_0^2}{n+\frac{1}{2}-\frac{|q|}{2q}\sqrt{1+\frac{4l'(l'+1)D_2}{a^2R_0^2}}}-\frac{1}{a^2}\left(n+\frac{1}{2}-\frac{|q|}{2q}\sqrt{1+\frac{4l'(l'+1)D_2}{a^2R_0^2}}\right)\right]^2, \quad (28)$$

where $l'(l'+1)=\left(2\tilde{n}+\sqrt{m^2+4M(\alpha+\beta)}+\frac{3}{2}\right)^2-4M\beta-\frac{1}{4}.$

For numerical solution of the energy equations (27) and (28) with parameter values $M=10\,fm^{-1}$, $R_0=7\,fm$, $a=0.5fm$ and $V_0=5fm^{-1}$ [54], we have approximately calculated the energy eigenvalues for the usual WS potential ($q=1$) plus the ring-shaped potential ($\alpha=1$ and $\beta=1$) and compare when the ring-shaped potential is zero ($\alpha=0$ and $\beta=0$). The results are shown in Table 1.

Further, we plot the energy behavior of the Dirac equation with WS potential plus RS potential versus the surface thickness $a$, the deformation parameter $q$ and nuclear radius $R_0$ for various values $n$, $\tilde{n}$ and $m$. This plot is shown in Figures 1 to 3. As shown in Figure 1, the energy becomes more negative as the surface thickness $a$ increases, i.e., the particle becomes more attractive by the potential (1). However, for Figure 2 and Figure 3. energy increases in the positive direction (less negative) with the increase of the deformation parameter $q$ and the nuclear radius $R_0$ for any given state $(n,\tilde{n},m)$, respectively, i.e., the particle becomes less attractive to potential (1).

Now we turn to calculate the corresponding radial part of the wave function. The first step, we find the weight function via (A11) as

$$\rho(s)=s^{2p_0}(1+qs)^{w_0}, \quad (29)$$

where

$$p_0=\frac{1}{\gamma}\sqrt{R_0^2\left[M^2-E^2-2(E+M)V_0\right]+l(l+1)(D_0+D_1+D_2)}>0, \quad (30a)$$

$$w_0=-\frac{1}{q}\sqrt{q^2\left(1+\frac{4l(l+1)D_2}{\gamma^2}\right)}. \quad (30b)$$

Hence, using Eq. (29), the first part of the radial wave function can be obtained by means of the relation (A13) in terms of the Jacobi polynomials as

$$y_n(s)\sim P_n^{(2p_0,w_0)}(1+2qs). \quad (31)$$

The second part of the radial wave function can be obtained via (A12) as

$$\phi(s)\sim s^{p_0}\left(1+qs\right)^{(1+w_0)/2}, \tag{32}$$

and hence the radial part of the wave function, $U_{nl}(s)=\phi(s)y_n(s)$ is

$$U_{nl}(r)=B_{nl}\left(e^{(r-R_0)/a}\right)^{p_0}\left(1+qe^{(r-R_0)/a}\right)^{(1+w_0)/2}P_n^{(2p_0,w_0)}\left(1+2qe^{(r-R_0)/a}\right), \tag{33}$$

where $B_{nl}$ is the normalization constant and we have used the definition of the Jacobi polynomials given by [55]

$$P_n^{(a,b)}(y)=\frac{(-1)^n}{n!2^2}(1-y)^{-a}(1+y)^{-b}\frac{d^n}{dy^n}\left[(1-y)^{a+n}(1+y)^{b+n}\right].$$

To compute the normalization constant $B_{nl}$, it is easy to show with the use of $R_{nl}(r)=r^{-1}U_{nl}(r)$, that

$$\int_0^\infty\left|R_{nl}(r)\right|^2r^2dr=\int_0^\infty\left|U_{nl}(r)\right|^2dr=\int_0^\infty\left|U_{nl}(s)\right|^2\frac{ads}{s}=1, \tag{34}$$

where we have used the substitution $s=e^{(r-R_0)/a}$. In our case, with the aid of (A15), the Jacobi polynomials can be expressed in terms of the hypergeometric function as [55]

$$P_n^{(2p_0,w_0)}\left(1+2qe^{(r-R_0)/a}\right)=\frac{\Gamma\left(n+2p_0+1\right)}{n!\Gamma\left(2p_0+1\right)}\,{}_2F_1\left(-n,2p_0+w_0+n+1;1+2p_0;qe^{(r-R_0)/a}\right). \tag{35}$$

Finally, combining Eqs. (14), (21) and (33), the total upper-component of the wave function (11) becomes

$$\varphi(\vec{r})=\varphi_{nlm}(r,\theta,\phi)=N_{nlm}\frac{1}{\sqrt{2\pi}}\cos\theta\left(\sin\theta\right)^{\sqrt{m^2+2(E+M)(\alpha+\beta)}}$$

$$\times P_{\tilde{n}}^{\left(1/2,\sqrt{m^2+2(E+M)(\alpha+\beta)}\right)}\left(1-2\cos^2\theta\right)e^{\pm im\phi}\left(e^{(r-R_0)/a}\right)^{p_0}$$

$$\times\left(1+qe^{(r-R_0)/a}\right)^{(1+w_0)/2}P_n^{(2p_0,w_0)}\left(1+2qe^{(r-R_0)/a}\right), \tag{36}$$

where $m=0,1,2,\cdots$, $n=0,1,2,\cdots$, and recalling that

$$l=\sqrt{\left(2\tilde{n}+\sqrt{m^2+2(E+M)\left(\alpha+\beta\right)}+\frac{3}{2}\right)^2-2(E+M)\beta}-\frac{1}{2},\ l=0,1,2,\cdots.$$

The lower-component of the wave function (5) can be found by means of Eq. (8b) as

$$\chi(\vec{r})=\chi_{nlm}(r,\theta,\phi)=N_{nlm}\frac{1}{\sqrt{2\pi}}\frac{\vec{\sigma}\cdot\vec{p}}{E+M}\cos\theta\left(\sin\theta\right)^{\sqrt{m^2+2(E+M)(\alpha+\beta)}}$$

$$\times P_{\tilde{n}}^{\left(1/2,\sqrt{m^2+2(E+M)(\alpha+\beta)}\right)}\left(1-2\cos^2\theta\right)e^{\pm im\phi}\left(e^{(r-R_0)/a}\right)^{p_0}$$

$$\times\left(1+qe^{(r-R_0)/a}\right)^{(1+w_0)/2} P_n^{(2p_0,w_0)}\left(1+2qe^{(r-R_0)/a}\right),\quad E\neq -M \tag{37}$$

For the case of the ring-shaped Hulthen potential, we make the following simple changes: $q=-e^{R_0/a}$ and $V_0=-V_0'$ in the expressions (27), (30), (33), (36) and (37).

## 4. Final remarks and conclusion

In this work, we have investigated the approximate bound state solutions of the Dirac equation with the $q$-deformed WS plus a new ring-shaped potential for any orbital $l$ quantum numbers. By making an appropriate approximation to deal with the centrifugal potential term, we have obtained the energy eigenvalue equation and the unnormalized two spinor components of the wave function $\varphi(\vec{r})$ and $\chi(\vec{r})$ expressed in terms of the Jacobi polynomials. This problem is solved within the shortcut of the NU method introduced recently in [56]. The relativistic solution can be reduced into the Schrödinger solution under the nonrelativistic limit, to the Hulthen solution and to ring-shaped usual WS potential with ($q=1$).

## Appendix A: A Shortcut of the NU Method

The NU method is used to solve second order differential equations with an appropriate coordinate transformation $s = s(r)$ [33]

$$\psi_n''(s) + \frac{\tilde{\tau}(s)}{\sigma(s)}\psi_n'(s) + \frac{\tilde{\sigma}(s)}{\sigma^2(s)}\psi_n(s) = 0, \tag{A1}$$

where $\sigma(s)$ and $\tilde{\sigma}(s)$ are polynomials, at most of second degree, and $\tilde{\tau}(s)$ is a first-degree polynomial. To make the application of the NU method simpler and direct without need to check the validity of solution. We present a shortcut for the method. So, at first we write the general form of the Schrödinger-like equation (A1) in a more general form applicable to any potential as follows [56-59]

$$\psi_n''(s)+\frac{(c_1-c_2s)}{s(1-c_3s)}\psi_n'(s)+\frac{(-As^2+Bs-C)}{s^2(1-c_3s)^2}\psi_n(s)=0, \tag{A2}$$

satisfying the wave functions

$$\psi_n(s)=\phi(s)y_n(s). \tag{A3}$$

Comparing (A2) with its counterpart (A1), we obtain the following identifications:

$$\tilde{\tau}(s)=c_1-c_2s, \quad \sigma(s)=s(1-c_3s), \quad \tilde{\sigma}(s)=-As^2+Bs-C, \tag{A4}$$

Following the NU method [33], we obtain the following necessary parameters [56],

(i) Relevant constant:

$$c_4=\frac{1}{2}(1-c_1), \qquad c_5=\frac{1}{2}(c_2-2c_3),$$

$$c_6=c_5^2+A, \qquad c_7=2c_4c_5-B,$$

$$c_8=c_4^2+C, \qquad c_9=c_3(c_7+c_3c_8)+c_6,$$

$$c_{10}=c_1+2c_4+2\sqrt{c_8}-1>-1, \qquad c_{11}=1-c_1-2c_4+\frac{2}{c_3}\sqrt{c_9}>-1,\ c_3\neq 0,$$

$$c_{12}=c_4+\sqrt{c_8}>0, \qquad c_{13}=-c_4+\frac{1}{c_3}(\sqrt{c_3}-c_5)>0,\ c_3\neq 0. \tag{A5}$$

(ii) Essential polynomial functions:

$$\pi(s)=c_4+c_5s-\left[\left(\sqrt{c_9}+c_3\sqrt{c_8}\right)s-\sqrt{c_8}\right], \tag{A6}$$

$$k=-(c_7+2c_3c_8)-2\sqrt{c_8c_9}, \tag{A7}$$

$$\tau(s)=c_1+2c_4-(c_2-2c_5)s-2\left[\left(\sqrt{c_9}+c_3\sqrt{c_8}\right)s-\sqrt{c_8}\right], \tag{A8}$$

$$\tau'(s)=-2c_3-2\left(\sqrt{c_9}+c_3\sqrt{c_8}\right)<0. \tag{A9}$$

(iii) Energy equation:

$$c_2n-(2n+1)c_5+(2n+1)\left(\sqrt{c_9}+c_3\sqrt{c_8}\right)+n(n-1)c_3+c_7+2c_3c_8+2\sqrt{c_8c_9}=0. \tag{A10}$$

(iv) Wave functions

$$\rho(s)=s^{c_{10}}(1-c_3s)^{c_{11}}, \tag{A11}$$

$$\phi(s)=s^{c_{12}}(1-c_3s)^{c_{13}},\ c_{12}>0,\ c_{13}>0, \tag{A12}$$

$$y_n(s)=P_n^{(c_{10},c_{11})}(1-2c_3s),\ c_{10}>-1,\ c_{11}>-1, \tag{A13}$$

$$\psi_{n\kappa}(s)=N_{n\kappa}s^{c_{12}}(1-c_3s)^{c_{13}}P_n^{(c_{10},c_{11})}(1-2c_3s). \tag{A14}$$

where $P_n^{(\mu,\nu)}(x)$, $\mu > -1$, $\nu > -1$, and $x \in [-1,1]$ are Jacobi polynomials with

$$P_n^{(a_0,b_0)}(1-2s) = \frac{(a_0+1)_n}{n!}\,{}_2F_1(-n, 1+a_0+b_0+n; a_0+1; s), \tag{A15}$$

and $N_{n\kappa}$ is a normalization constant. Also, the above wave functions can be expressed in terms of the hypergeometric function as

$$\psi_{n\kappa}(s) = N_{n\kappa} s^{c_{12}} (1-c_3 s)^{c_{13}}\,{}_2F_1(-n, 1+c_{10}+c_{11}+n; c_{10}+1; c_3 s) \tag{A16}$$

where $c_{12} > 0$, $c_{13} > 0$ and $s \in [0, 1/c_3]$, $c_3 \neq 0$.

**Appendix B: Approximation to the Strong Singular Orbital Centrifugal Term**

Here we make a new approximation to deal with the strong singular centrifugal potential given in Eq. (12a). The centrifugal term is expanded around $r = R_0$ or $x = 0$ in a series of powers of $x = (r - R_0)/R_0 \in (-1, \infty)$ and $\gamma = R_0 / a$ as

$$V_l(r) = \frac{\lambda}{r^2} = \frac{\lambda}{R_0^2(1+x)^2} = \frac{\lambda}{R_0^2}(1 - 2x + 3x^2 - ...), \quad x \ll 1, \tag{B1}$$

where $\lambda = l(l+1)$. The above centrifugal potential (B1) can be replaced by the form formally homogeneous to the original $q$-deformed Woods-Saxon potential to keep the factorization of the corresponding Schrödinger-like equation. Thus, we take the centrifugal potential in the form

$$\tilde{V}_l(r) = \frac{\lambda}{R_0^2}\left[D_0 + D_1 \frac{1}{1+qe^{\gamma x}} + D_2 \frac{1}{(1+qe^{\gamma x})^2}\right], \quad x \ll 1/\gamma, \quad \gamma x \ll 1 \tag{B2}$$

where $\gamma = R_0 / a$ and $D_i$ are the parameter of coefficients ($(i = 0,1,2)$). After making a Taylor expansion to (B2) up to the second order term, $x^2$, and then comparing equal powers with (B1), we can readily determine $D_i$ $(i = 0,1,2)$ parameters of the generalized Woods-Saxon potential as function of the specific potential parameters $R_0$, $\gamma$ and $q$ as follows:

$$D_0 = 1 - \left(3 + \frac{2}{q} - \frac{1}{q^2}\right)\frac{1}{\gamma} + 3\left(1 + \frac{2}{q} + \frac{1}{q^2}\right)\frac{1}{\gamma^2}, \tag{B3}$$

$$D_1 = 2\left(3 + 2q - \frac{1}{q^2}\right)\frac{1}{\gamma} - 6\left(3 + q + \frac{3}{q} + \frac{1}{q^2}\right)\frac{1}{\gamma^2}, \tag{B4}$$

And

$$D_2 = -\left(2q + 3q^2 - \frac{2}{q} - \frac{1}{q^2}\right)\frac{1}{\gamma} + 3\left(6 + 4q + q^2 + \frac{4}{q} + \frac{1}{q^2}\right)\frac{1}{\gamma^2}. \quad \text{(B5)}$$

When $q = 1$, we obtain an approximation for the centrifugal potential in case of usual WS potential as

$$D_0 = 4\left(\frac{3}{\gamma^2} - \frac{1}{\gamma}\right) + 1, \quad D_1 = 8\left(\frac{1}{\gamma} - \frac{6}{\gamma^2}\right), \quad D_2 = 2\left(\frac{24}{\gamma^2} - \frac{1}{\gamma}\right). \quad \text{(B6)}$$

**Table 1.** The bound-state energy eigenvalues of the usual WS potential plus RS potential for various values of $n$, $\tilde{n}$ and $m$ quantum numbers.

| $n$ | $\tilde{n}$ | $m$ | $\alpha = \beta = 1$ | | $\alpha = \beta = 0$ | |
|---|---|---|---|---|---|---|
| | | | $E_{n,\tilde{n},m}$ relativistic | $E_{n,\tilde{n},m}$ non-relativistic | $E_{n,\tilde{n},m}$ relativistic | $E_{n,\tilde{n},m}$ non-relativistic |
| 1 | 0 | 0 | −8.181582591 | −15.72190029 | −8.087401684 | −13.64437737 |
| 1 | 0 | 1 | −8.199178439 | −16.35979304 | −8.245532350 | −15.00275622 |
| 1 | 1 | 0 | −8.390837180 | −23.88322236 | −8.449864847 | −17.25283431 |
| 1 | 1 | 1 | −8.930386266 | −25.42691358 | −8.676477943 | −20.76166288 |
| 2 | 0 | 0 | −5.288469326 | −5.163723495 | −4.511739044 | −5.131551195 |
| 2 | 0 | 1 | −5.32440643 | −5.175961484 | −4.671607147 | −5.151060868 |
| 2 | 1 | 0 | −5.923074363 | −5.351169122 | −4.890413294 | −5.194344782 |
| 2 | 1 | 1 | −5.962042737 | −5.388986308 | −5.153319654 | −5.274965974 |
| 2 | 2 | 0 | −6.553753850 | −5.829767768 | −5.448307237 | −5.409155814 |
| 2 | 2 | 1 | −6.595079988 | −5.915307507 | −5.765975978 | −5.617085700 |
| 3 | 0 | 0 | −2.480509032 | −5.593696192 | −1.844569412 | −5.809477801 |
| 3 | 0 | 1 | −2.501689705 | −5.537243884 | −1.938721382 | −5.662367950 |
| 3 | 1 | 0 | −2.967053503 | −5.073595802 | −2.069933874 | −5.464359829 |
| 3 | 1 | 1 | −2.990765343 | −5.007785930 | −2.231472167 | −5.231175524 |

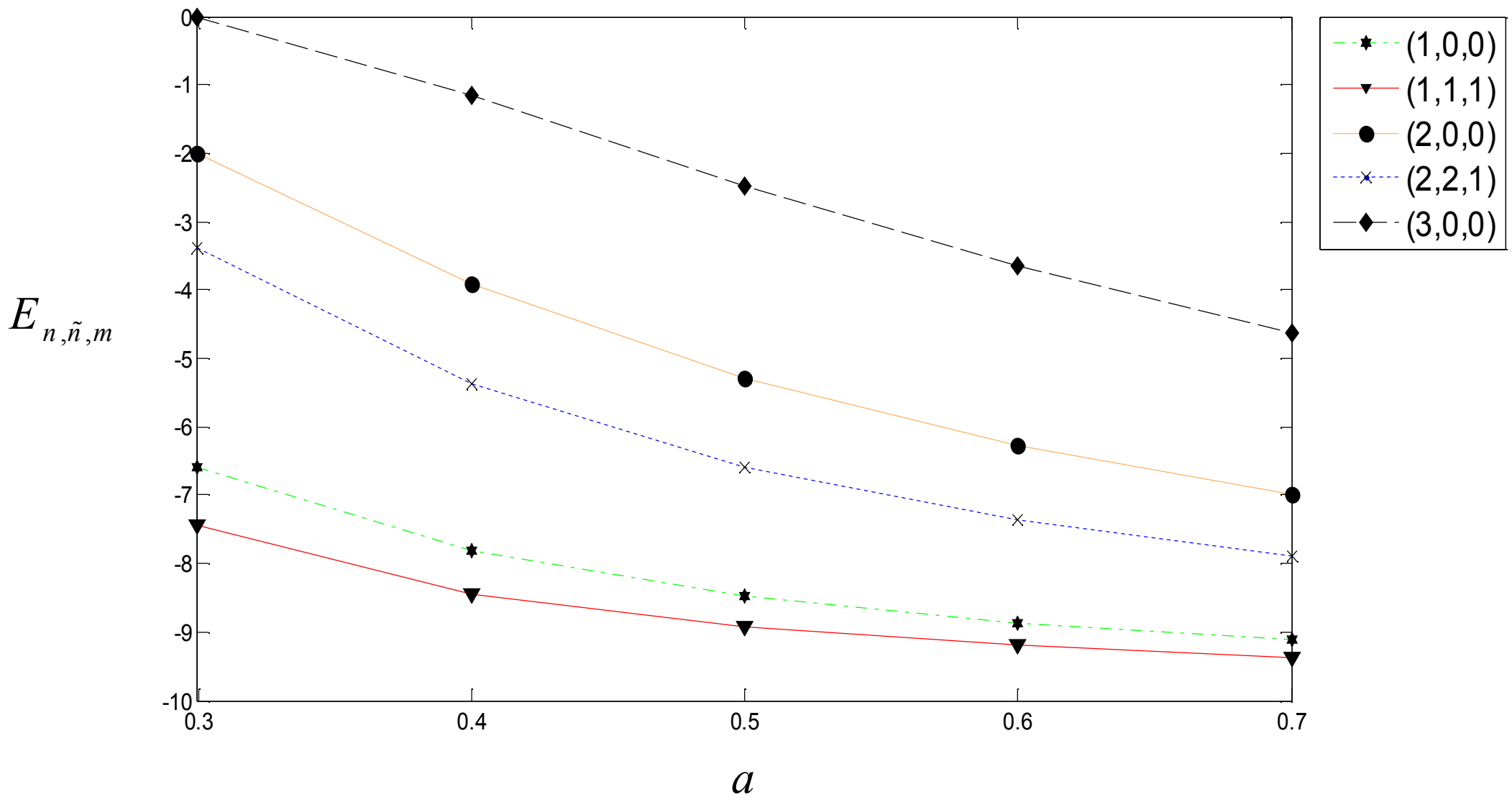

**Fig. 1:** Energy behavior of the Dirac equation with WS potential plus RS potential versus the diffuseness of the nuclear surface $a$ for various $n$, $\tilde{n}$ and $m$, respectively.

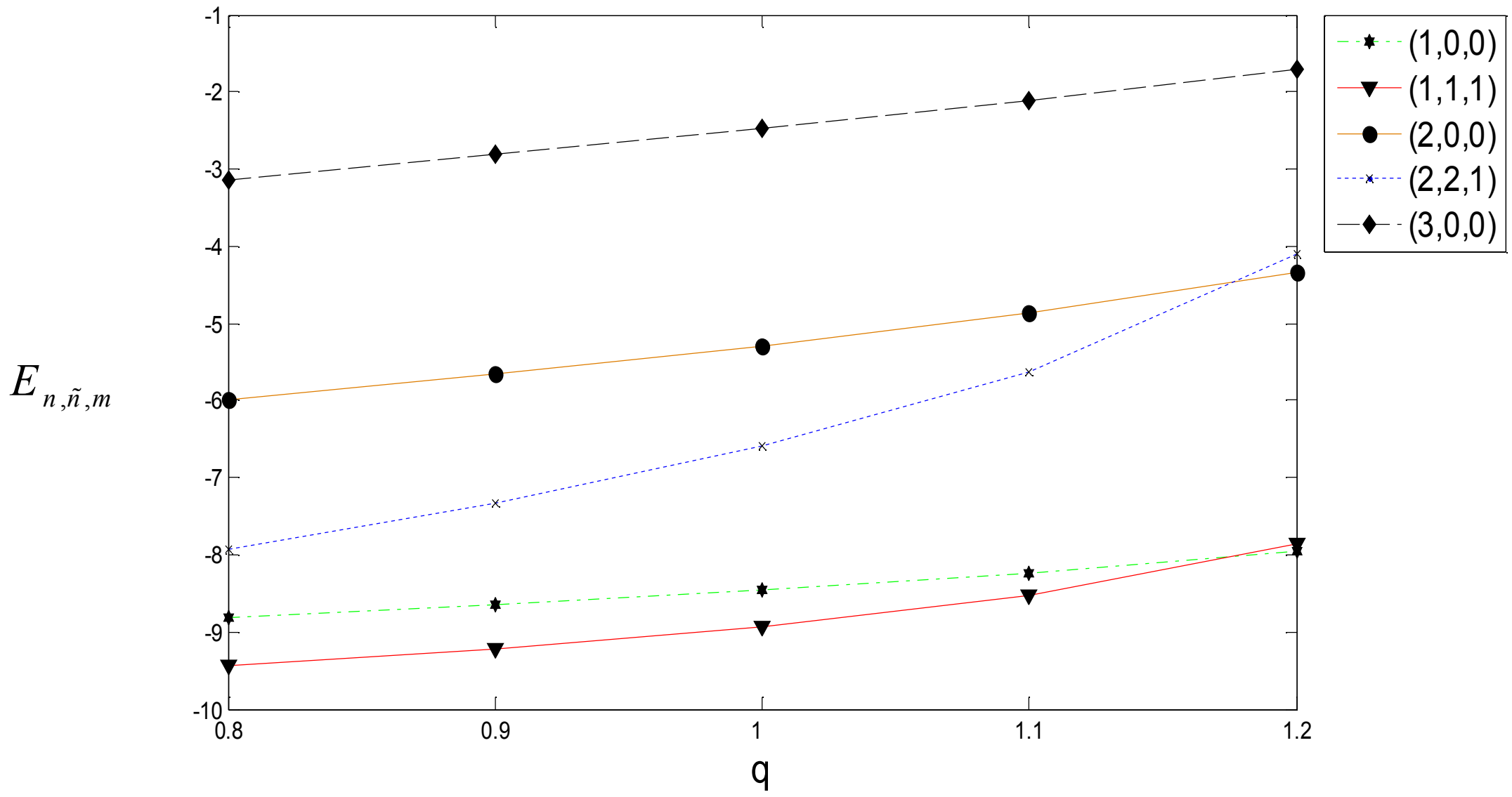

**Fig. 2:** Energy behavior of the Dirac equation with WS potential plus RS potential versus deformation parameter $q$ for various $n$, $\tilde{n}$ and $m$ .

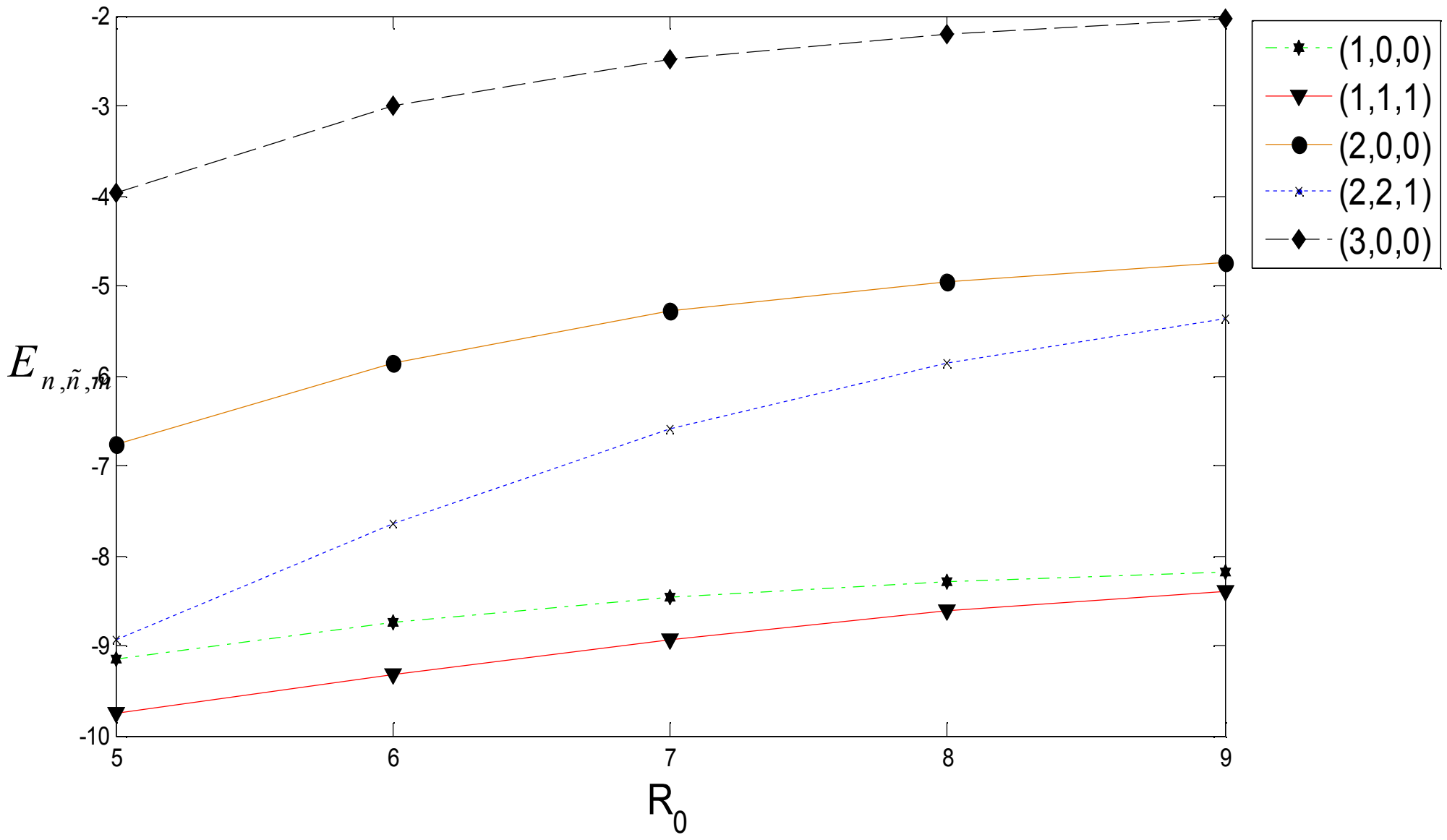

**Fig. 3:** Energy behavior of the Dirac equation with WS potential plus RS potential versus width of the potential $R_0$ for various $n$, $\tilde{n}$ and $m$, respectively.